\PassOptionsToPackage{dvipsnames}{xcolor}

\documentclass[sigconf]{acmart}
\usepackage{tikz,pgfplots,pgfplotstable,multicol}
\AtBeginDocument{%
  }

\setcopyright{acmlicensed}
\copyrightyear{2018}
\acmYear{2018}
\acmDOI{XXXXXXX.XXXXXXX}
\acmConference[Conference acronym 'XX]{Make sure to enter the correct
  conference title from your rights confirmation email}{June 03--05,
  2018}{Woodstock, NY}
\acmISBN{978-1-4503-XXXX-X/2018/06}

\title{Co-Lecturing With the DED}

\subtitle{Explaining Circuit Design via the Draw Encode Display Loop}

\newtoggle{BLIND}
\togglefalse{BLIND}
\newcommand{\blinding}[1]{\iftoggle{BLIND}{\textbf{REDACTED}}{#1}}

\iftoggle{BLIND}{
\author{Anonymous}
\email{anon@anon.com}
\orcid{}
\affiliation{%
  \institution{University of Anonymous}
  \streetaddress{}
  \city{Anonymous City}
  \country{Anonymous Country}
  \postcode{}}
}{
\author{Alasdair Lambert}
\email{alasdair.lambert@strath.ac.uk}
\orcid{0000-0002-9762-2193}
\affiliation{%
 \institution{University of Strathclyde}
 \streetaddress{26 Richmond Street}
 \city{Glasgow}
 \country{UK}
 \postcode{G1 1XH}
}

\author{Guillaume Allais}
\email{guillaume.allais@strath.ac.uk}
\orcid{0000-0002-4091-657X}
\affiliation{%
 \institution{University of Strathclyde}
 \streetaddress{26 Richmond Street}
 \city{Glasgow}
 \country{UK}
 \postcode{G1 1XH}
}

\author{Conor Mc Bride}
\email{conor.mcbride@strath.ac.uk}
\orcid{0000-0003-1487-0886}
\affiliation{%
 \institution{University of Strathclyde}
 \streetaddress{26 Richmond Street}
 \city{Glasgow}
 \country{UK}
 \postcode{G1 1XH}
}
}



\usepackage[dvipsnames]{xcolor}
\usepackage{lineno}
\newcommand{\markblankline}{\par\mbox{}\par}
\usepackage{changepage}

\usepackage{tikz}
\usepackage{pgfplots}
\pgfplotsset{compat=1.18}

\usepackage{cleveref}

\usepackage{todonotes}
\setuptodonotes{inline}

\newcommand{\paragraphy}[1]{\textit{#1.}}

\newcommand{\syrupKeyword}[1]{{\tt\textbf{#1}}}
\newcommand{\syrupType}[1]{{\tt\color{OliveGreen}{<#1>}}}

\newcommand{\syrupBit}{\syrupType{Bit}}
\newcommand{\syrupCable}[1]{\syrupKeyword{[}#1\syrupKeyword{]}}

\newcommand{\syrupVarName}[1]{{\tt#1}}

\newcommand{\syrupTypeDecl}[2]
  {{\tt\syrupKeyword{type} \syrupType{#1} \syrupKeyword{=} #2}}

\newcommand{\syrupFunName}[1]{{\tt\color{blue}{#1}}}

\newcommand{\syrupFunCall}[2]
  {\syrupFunName{#1}\syrupKeyword{(}#2\syrupKeyword{)}}

\newcommand{\syrupFunDecl}[3]
  {\syrupFunCall{#1}{#2} \syrupKeyword{->} #3}

\newcommand{\syrupFunDefnWhereLine}[4]
  {\syrupFunCall{#1}{#2} \syrupKeyword{=} #3 \syrupKeyword{where}
      \begin{adjustwidth}{1em}{0em}
        #4
      \end{adjustwidth}}

\newcommand{\syrupFunDefnWhereBlock}[4]
  {\syrupFunCall{#1}{#2}
    \begin{adjustwidth}{1em}{0em}
      \syrupKeyword{=} #3
      \syrupKeyword{where}
      \begin{adjustwidth}{1em}{0em}
        #4
      \end{adjustwidth}
    \end{adjustwidth}}

\newcommand{\syrupFunDefn}[3]
  {\syrupFunCall{#1}{#2} \syrupKeyword{=} #3}

\newcommand{\syrupEquation}[3]
  {#1 \syrupKeyword{=} \syrupFunCall{#2}{#3}}

\definecolor{Firebrick}{RGB}{178,34,34}
\newcommand{\syrupComment}[1]
  {{\tt\color{Firebrick}{-{}- #1}}}

\makeatletter\newenvironment{syrup}%
  {\raggedright%
   \begin{linenumbers}%
   \setlength{\parindent}{0pt}%
   \tt}
  {\end{linenumbers}%
   \resetlinenumber}

\begin{document}

\begin{abstract}

When representing digital circuits, 2 dimensional hand drawings free
us from the linear structure of hardware description languages,
enabling intuitive reasoning and making structure explicit.
However these drawings are imprecise and inert: they do not enforce
that the circuits are well defined and cannot be tested.

We want both intuitive visual representations
and well defined testable ones
but students can struggle to link one ot the other.

To bridge this gap we present the Draw Encode Display Loop (DED), a
Co-Lecturing dynamic which equips students with a systematic
method to tackle natural language specifications:

\begin{enumerate}
 \item \textbf{Draw} visually informative intermediate representations
 (truth tables, characteristic tables) to generate a structured
 circuit diagram.
 \item \textbf{Encode} the diagram by labelling
 inputs, outputs and intermediate values which can be directly
 converted to code.
 \item \textbf{Display} the code using an in-house
 diagrammatic renderer.
\end{enumerate}

This is supported by Syrup, an education-focused hardware description
language which allows students to define, experiment with and display
their own circuits. We support these approaches with survey data
gathered from two cohorts of students.

\end{abstract}

\maketitle

\section{Introduction}

By focusing on high-level understanding, a visual representation of a
complex and technical topic can reduce the cognitive load that would
be associated with a textual implementation.
Clearly this has pedagogical value as high level understanding provides
a foundation on which to build detailed knowledge~\cite{notional}.
But visualisations are a double edged sword for Computer Science educators:
heavy use of visual strategies can mean that
students are less confident when presented with text-based
programming languages~\cite{transition2text}.

\begin{figure}[h]
\includegraphics[width=.375\textwidth]{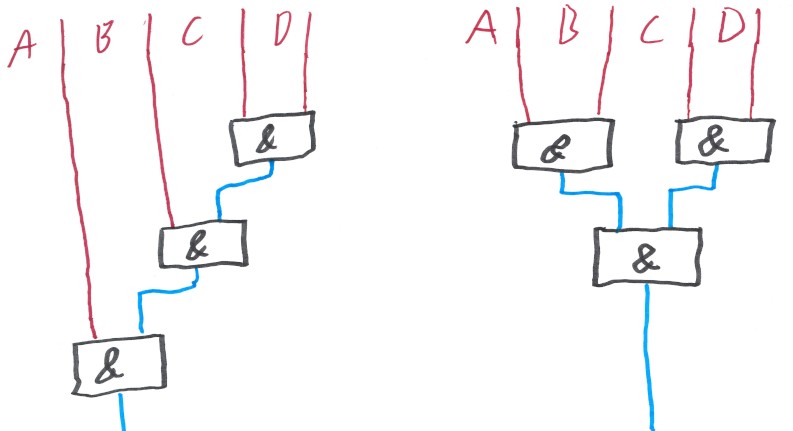}
\caption{Sequential and Parallel And for 4 inputs as diagrams}
\label{fig:diagramand4}
\end{figure}

When teaching digital logic to first year undergraduate
students we like to
first draw a rich and colourful diagram showing the problem's structure,
and then translate it into a formal description language which allows us
to check our solution has the correct behaviour.
This is because small digital circuits have a clear flow of
information from inputs, through wires, into logic gates, and towards
outputs.
Correspondingly, diagrams are a natural visualisation tool for such
circuits with varied colours making the information flow explicit and
a well chosen layout highlighting their structural properties.
To wit, \cref{fig:diagramand4} shows how distinct the diagrams for the
sequential and parallel versions of a circuit taking the conjunction
of its 4 inputs look.

We also use textual hardware desription languages because they can
easily be processed by computers.
They enable users to make sure the wiring is correct, and
interactively simulate these circuits' behaviour.
These languages may however obfuscate the physical structure of the
circuits and make it harder to distinguish between functionally
equivalent but structurally different components as demonstrated
by \cref{fig:syrupand4}.
This figure shows how the hardware description implementations%
\footnote
  {We will explain the syntax of Syrup definitions in \cref{sec:syruplang};
  for now it is enough to play `spot the differences' between the two
  definitions.}
for the sequential and parallel circuits pictured above
only differ in how parentheses are arranged.
A difference that could easily be missed in a much bigger description.

\begin{figure}[h]
\begin{syrup}
\syrupFunDecl{and4seq}
  {\syrupBit\syrupKeyword{,} \syrupBit\syrupKeyword{,} \syrupBit\syrupKeyword{,} \syrupBit}
  {\syrupBit}

\syrupFunDefn{and4seq}
  {A\syrupKeyword{,} B\syrupKeyword{,} C\syrupKeyword{,} D}
  {A \syrupFunName{\&} \syrupKeyword{(}B \syrupFunName{\&} \syrupKeyword{(}C \syrupFunName{\&} D\syrupKeyword{)}\syrupKeyword{)}}

\markblankline

\syrupFunDecl{and4par}
  {\syrupBit\syrupKeyword{,} \syrupBit\syrupKeyword{,} \syrupBit\syrupKeyword{,} \syrupBit}
  {\syrupBit}

\syrupFunDefn{and4par}
  {A\syrupKeyword{,} B\syrupKeyword{,} C\syrupKeyword{,} D}
  {\syrupKeyword{(}A \syrupFunName{\&} B\syrupKeyword{)} \syrupFunName{\&} \syrupKeyword{(}C \syrupFunName{\&} D\syrupKeyword{)}}
\end{syrup}
\caption{Sequential and Parallel And for 4 inputs in Syrup}
\label{fig:syrupand4}
\end{figure}

But by using this visual-first approach
we are cut by the aforementioned double edged sword:
we present a high-level visualisation of circuits to solve
a given problem
and we need our students to convert these visualisations
to a hardware description language.

To resolve this we present the Draw-Encode-Display (DED) loop, a
co-lecturing dynamic for delivering circuit design lectures which
highlights the connection between visual circuit representations and
their formal implementation.
Here one lecturer solves a problem using pen and paper (Draw),
discussing how to approach and resolve problems with the available
tools, before handing the diagram over.
The other lecturer can then focus on formalising the diagram (Encode).
Finally, they close the loop by checking that the code corresponds to
the hand drawn solution by using the display feature (Display).

The DED loop has been used for two years during semester one of
\blinding{CS106 Computer Systems and Organisation}.
This is a core first year module with approximately 180 students per
year, the course covers digital logic and basic instruction set
architecture

By surveying two cohorts of students we present
preliminary data addressing the following research questions.

\begin{itemize}
\item[\textbf{RQ1}] Do students find the DED loop helpful?
\item[\textbf{RQ2}] Do students prefer the DED loop to a code-only approach?
\item[\textbf{RQ3}] Did students find Co-Lecturing helpful?
\end{itemize}

\section{Related Work}

The DED loop is a co-lecturing strategy involving visual
representations of computational objects and live programming.

\paragraphy{Visual Representations of Programs}
Block-based programming, where learners build programs by composing
compatible blocks which perform small actions, is an established tool for
beginner programmers~\cite{schoolBlock,blockCS,blockvstext,transition2text,block2textMeta,blockornot}.
These tools, due to their visual nature, allow beginners to write code
more easily than with text-based programming
languages~\cite{schoolBlock,blockvstext} and have a small but positive
impact on academic outcomes~\cite{blockornot,blockvstext}.
These language are engaging and make composition of effects
easier~\cite{schoolBlock} but they suffer from a perceived
inauthenticity as they are not industrial strength~\cite{schoolBlock}
and can cause frustration when moving to text-based
languages~\cite{transition2text}.
%
Circuit diagrams function in much the same way: small components are
wired together to create larger circuits.
We can therefore expect the same benefits and drawbacks when using hand
drawn circuits, however with a lack of machine verified feedback for
students.

\paragraphy{Live Programming}
Our work relies heavily on the Cognitive Apprenticheship (CA)
model~\cite{collins1991cognitive} combined with live
programming~\cite{Knobelsdorf2014,Yu2024,paxton2002,Selvaraj2021,raj2018role}.
The CA model focuses on making thinking explicit by having educators
talk through their thought process as they solve
problems~\cite{collins1991cognitive}.
This can help learners understand topics that can otherwise seem
theoretical, or inaccessible~\cite{Knobelsdorf2014}.
It has a natural pairing with live programming, the engaging practice
of interactively building a program from a specification for the
benefit of the audience.
However live programming is far from a panacea for programming
education~\cite{Selvaraj2021} and has been shown to be similar in terms of student
outcomes to static delivery~\cite{liveEval}.
Similarly, the CA approach has proven benefits in CS education however
it struggles to scale well in the scaffolding and fading
stages~\cite{caReview}.

\paragraphy{Co-Lecturing}
Co-lecturing~\cite{Fagan2025} is a form of team teaching where
multiple lecturers present each lecture.
Team teaching has an established track record at
all education levels~\cite{Baeten2014,Zehetmeier2018,Liebel2017,coTA}.
In primary education it has been shown to have a positive impact on
educational attainment~\cite{Baeten2014}.
However at tertiary education it is less well explored and often
focuses on groups of teaching assistants~\cite{coTA}.
Co-Lecturing has been shown to be useful tool for training~\cite{Liebel2017},
where new lecturers can join existing teams, shoring up resilience to
staffing changes while acting as a critical friend~\cite{costa1993through}
for module enhancement.
Co-Lecturing can be deployed for particular
examples~\cite{Lambert2025} but has been shown to be particularly
successful when combined with the CA model for live
programming~\cite{Fagan2025}.

The DED loop uses aspects of visual programming and live programming,
delivered via co-lecturing with an emphasis on CA delivery.
Taken as a group we believe these techniques harmonise to provide a
holistic delivery method.
Pen and paper drawing are used to exploit the benefits of visual
programming such as lower cognitive load by reduced syntactic
noise and ease of compostition.
A hardware description language is then taught via co-lecturing and
live programming.
The final stages, encode and display are intended to show students
how to convert from diagrams to code helping to avoid common problems
with visual languages.
\section{Approach}

The DED loop has been used for two years during semester one of
\blinding{CS106 Computer Systems and Organisation}.
As the class is only taken by Computer Science or Software Engineering
Students, we do not use standard Electrical Engineering notations for
logic gates.
This reduces the time students spend learning notation, but also means
lectures focus on the logic and meaning of circuits, as well as
building large complex systems from small well understood ones.

Students use Syrup%
\footnote{
  Syrup is freely available at
  \blinding{\url{https://github.com/pigworker/Syrup}},
  and an interactive tutorial page can be found at
  \blinding{\url{https://personal.cis.strath.ac.uk/guillaume.allais/syrup/}}.
},
a bespoke hardware description language written in Haskell, to define
circuits that we describe in~\cref{sec:syruplang}.

Lectures are delivered via co-lecturing with a heavy focus on the
CA delivery model.
Each lecture is presented as a series of problems to be solved which
are tailored to illustrate the topics of the week.
Lectures are split between the two lecturers, one introduces the ideas
or problems for the lecture and then works through them using pen and
paper.
Colours, selected from a colourblind friendly palette to assure
accessibility, are used to indicate meaning e.g. place value.
Here the focus is to talk through the high level approach to a
problem, often starting with a english language problem description,
generating a truth table and then designing a circuit which implements
that behaviour.
This forms the Draw section of the lecture and aims to cement the strategy
for solving the problem in the minds of the students before any code is written.

Once a pen and paper solution is generated it is handed to the second
lecturer and the second part of the lecture begins, formalising the
pen and paper circuit via Syrup.
This is the Encode section of the lecture.
First they begin by listing the inputs and outputs, the bits that are
expected to be fed into the circuit and the ones that are expected as
output.
This forms the first line of the Syrup definition.
Intermediate wires are then given names that bear some meaning within
the circuit, as are the inputs and outputs.
This means that within the circuit definition the lecturer can show
that generating the code is simply a matter of chasing outputs
back through the diagram.
This shows students that if they have the diagram the code can be
generated systematically.
Once the circuit definition is complete the display feature is used to
show that the defined function is indeed the same as the pen and paper
solution.
This is the Display section of the lecture.
The ability to simulate the formalised circuit is also used in this stage
to make sure that the behaviour of the circuit is correct.

The DED loop is used during every lecture which involves building circuitry.
Roles for each lecturer are clearly defined, however both lectures can
interject or clarify during the others section.
Generally more time is spent on the Draw section as new topics are
introduced and the proposed solution needs to be thoroughly explained.
This allocation can change if new Syrup features need to be introduced,
or if class feedback indicates student need more support with Syrup.

\section{Syrup syntax}\label{sec:syruplang}

Syrup is a text-based statically typed programming language of circuit
descriptions built by connecting components using named wires.

\paragraphy{Types} Each wire has a type describing its structure: it
can either be carrying a single bit (\syrupBit), or be a cable packing
together $n$ wires, each with a given type $T_i$
(\syrupCable{$T_1\syrupKeyword{,} \dots\syrupKeyword{,} T_n$}).

\paragraphy{Declarations}
A component is declared by specifying its name and its interface
(i.e. the types of its inputs and outputs). For instance, in the code
below: line 1 declares a gate named \syrupFunName{not} with one input
and one output, both of type \syrupBit,
and line 2 and 3 respectively declare gates named \syrupFunName{and}
and \syrupFunName{or} both of which take two inputs and return one output,
all of type \syrupBit.

\begin{syrup}
\syrupFunDecl{not}{\syrupBit}{\syrupBit}

\syrupFunDecl{and}{\syrupBit\syrupKeyword{,} \syrupBit}{\syrupBit}

\syrupFunDecl{or}{\syrupBit\syrupKeyword{,} \syrupBit}{\syrupBit}
\end{syrup}

Without knowing the components' respective implementation, their
interfaces alone already tell us how they can be wired to other
components to build bigger circuits.

\paragraphy{Definitions}
A component is defined by giving a name to each of its input and using
various pre-existing components to produce its outputs.
For instance, in the code below: line 1 defines the \syrupFunName{not}
gate declared above by stating that its one input
is named \syrupVarName{X}, and that \syrupFunName{not} is defined by
duplicating this input, feeding it into both inputs of
a \syrupFunName{nand} gate\footnote{
The universal gate \syrupFunName{nand} comes pre-defined
in Syrup; everything else is ultimately defined in terms of it}
and immediately returning that result.
Once defined, a component can be freely reused in subsequent
definitions. For instance, line 2--3 give the standard definitions
of \syrupFunName{and} and \syrupFunName{or} in terms
of \syrupFunName{nand} and the \syrupFunName{not} gate we just
introduced.

\begin{syrup}
\syrupFunDefn{not}{\syrupVarName{X}}{\syrupFunCall{nand}{\syrupVarName{X}\syrupKeyword{,} \syrupVarName{X}}}

\syrupFunDefn{and}{\syrupVarName{X}\syrupKeyword{,} \syrupVarName{Y}}{\syrupFunCall{not}{\syrupFunCall{nand}{\syrupVarName{X}\syrupKeyword{,} \syrupVarName{Y}}}}

\syrupFunDefn{or}{\syrupVarName{X}\syrupKeyword{,} \syrupVarName{Y}}{\syrupFunCall{nand}{\syrupFunCall{not}{\syrupVarName{X}}\syrupKeyword{,} \syrupFunCall{not}{\syrupVarName{Y}}}}
\end{syrup}

Declarations are type checked to ensure our wiring respects
every component's interface.

\paragraphy{Where blocks} Larger definitions become difficult
to define without breaking up the logic. To ameleorate this Syrup
supports \syrupKeyword{where} blocks which allow students to name
intermediate outputs. This means that outputs can be implemented in
a step-by-step manner. See the definition of \syrupFunName{hadd}, a
half-adder which adds two bits together. Note that in the where block
meaningful names are given to the bit, representing their place value.

\begin{syrup}
  \syrupFunDecl{hadd}
    {\syrupBit\syrupKeyword{,} \syrupBit}
    {\syrupBit\syrupKeyword{,} \syrupBit}

  \syrupFunDefnWhereLine{hadd}
    {\syrupVarName{X}\syrupKeyword{,} \syrupVarName{Y}}
    {\syrupVarName{twos}\syrupKeyword{,} \syrupVarName{ones}}
    {\syrupEquation{\syrupVarName{twos}}{and}{\syrupVarName{X}\syrupKeyword{,} \syrupVarName{Y}}\\
     \syrupEquation{\syrupVarName{ones}}{xor}{\syrupVarName{X}\syrupKeyword{,} \syrupVarName{Y}}\\
    }
\end{syrup}

\begin{figure}[h]
  \includegraphics[width=.1\textwidth]{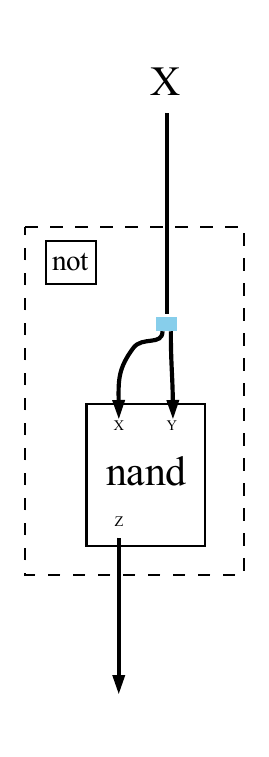}
  \hspace{.0275\textwidth}
  \includegraphics[width=.1\textwidth]{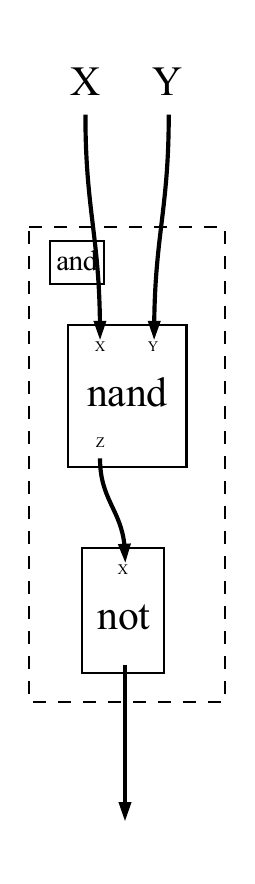}
  \hspace{.0275\textwidth}
  \includegraphics[width=.12\textwidth]{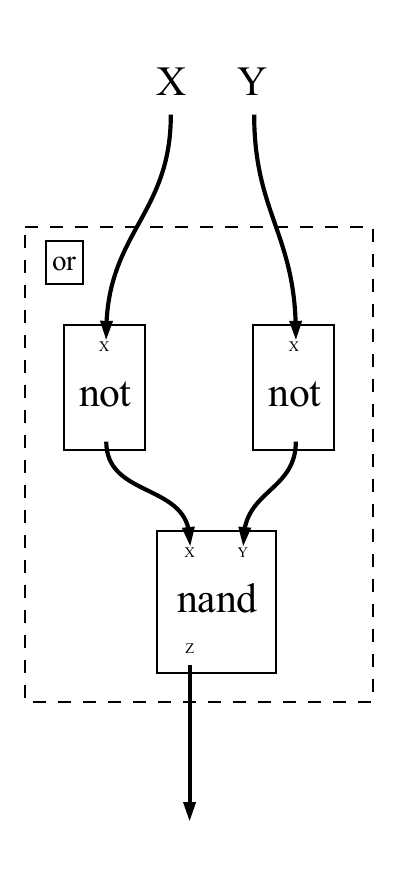}
  \caption{Syrup circuit diagrams of \syrupFunName{not}, \syrupFunName{and}, and \syrupFunName{or} defined in terms of \syrupFunName{nand}}
  \label{fig:displaylogic}
\end{figure}

\paragraphy{Experiments} Once we have defined components, we can
experiment with them. In this paper we are particularly interested in
the \syrupKeyword{display} function which allows us to visualise
(cf. \cref{fig:displaylogic}) the circuit diagrams corresponding to
the textual representations given above in order to check that our
code does indeed describe the circuits we expect.
Note that the box in \syrupFunName{not}'s diagram is an
explicit duplication node taking the signal named \syrupVarName{X}
and outputting two copies of it.

\begin{figure}[h]
  \includegraphics[width=.14\textwidth]{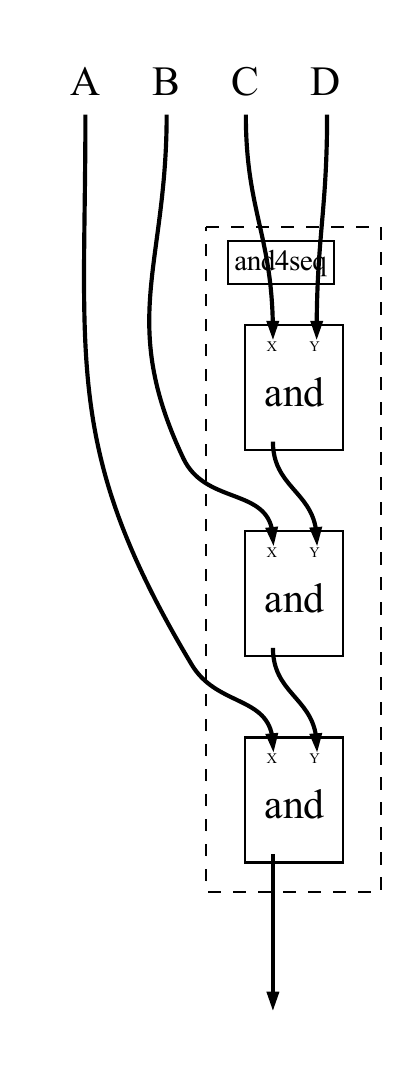}
  \hspace{.06\textwidth}
  \includegraphics[width=.14\textwidth]{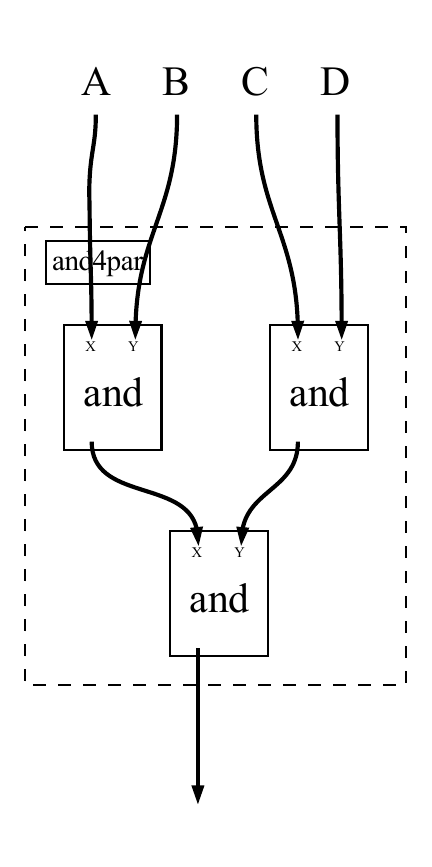}
  \caption{Variations on \syrupFunName{and4}}
  \label{fig:displayand4}
\end{figure}

By using \syrupKeyword{display} to visualise the sequential and
parallel variations on and4 shown in \cref{fig:syrupand4} we can see
(in \cref{fig:displayand4}) that we indeed obtain circuits that have
radically different structures.

\subsection{A DED-Easy Worked Example}

This example takes place at a point in the class where the students
have learned about a range of basic logic circuits, basic binary
arithmetic, half adders, and full adders.
For this example let us consider a ripple carry adder a circuit which
adds two cables of three bit numbers and a carry in together, returning
a cabled three bit number and a carry out.
We will refer to this circuit as rca3.
While lecturers often have names, we will refer to the lecturer
responsible for the draw section of the loop as the Architect and the
lecturer responsible for encoding the diagram as the Technician.
Before the lecture starts two projector screens are setup, the first
uses a document camera to display a blank sheet of paper and the other
a Syrup editor.
This means that students can see both the code and the written solution
at the same time.

\subsection{D for Draw}

The Architect begins with a problem statement and its context.
They recap that the class has covered how to add two bits together
using a half adder, and that they have improved on that with a full
adder to add three bits together.
They point out that the carry-in of a full adder makes it a composable
component so that carry-forwards can be accomodated.
The Architect then describes the problem, how do we add cables of bits
together?

\begin{figure}[h]
      \includegraphics[width=0.35\textwidth]{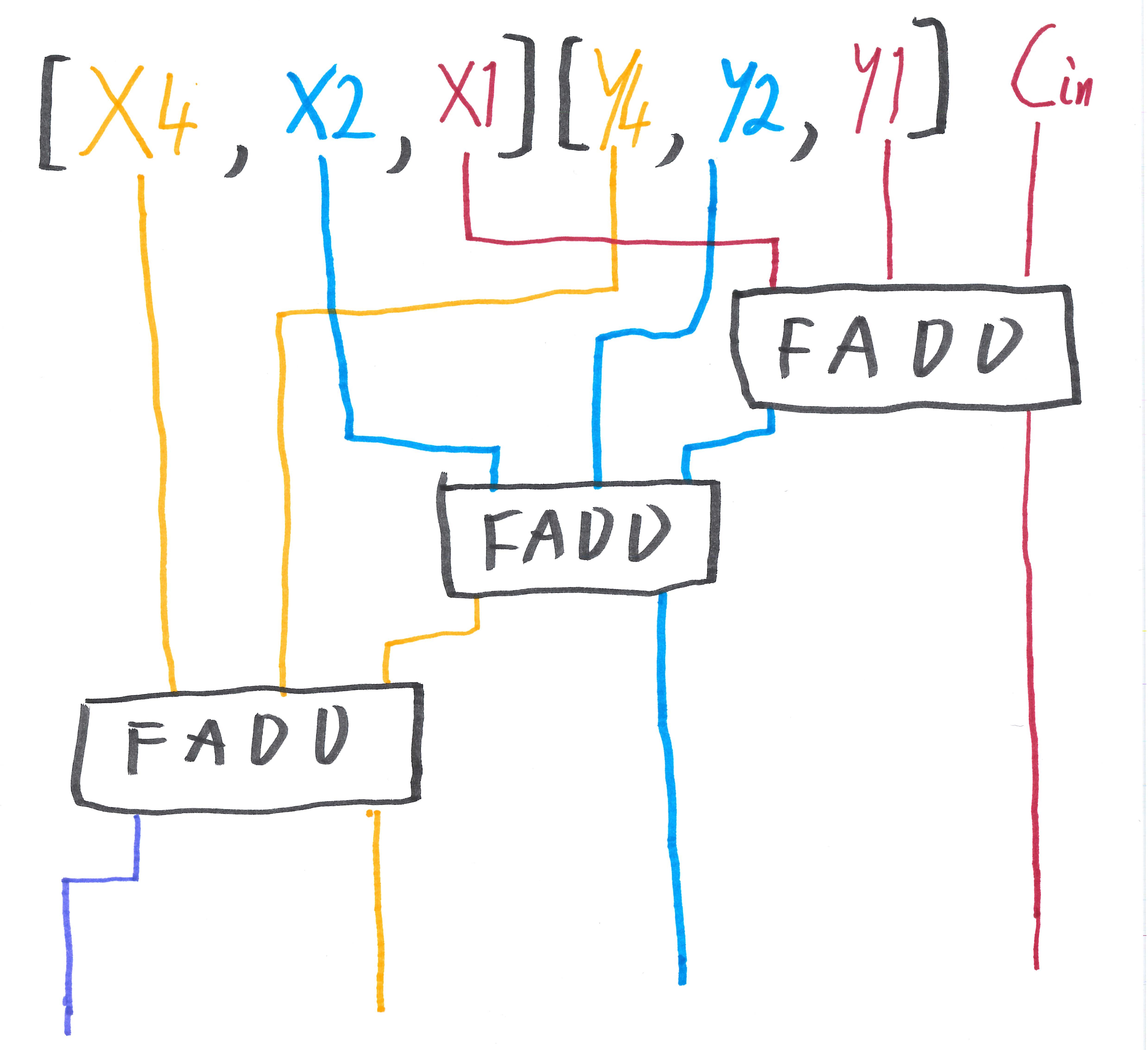}
      \caption{Solved RCA3}
      \label{fig:finDiag}
\end{figure}

The Architect sketches out the inputs to the circuit.
Multiple colours are used, the Architect asks the students why this is
the case?
In cases of stunned silence the Technician can chime in to ask ``is it
to do with place value?''.
Bits which have the same place value are given the same colour, the
Architect points out that this makes the problem simple, all we have
to do is deal with one place value at a time.
With this in mind the class is asked another question, ``Do we know of
a way to add three bits together?''.
The lecturers hope to hear the words ``full adder'', any bold and
correct student is rewarded with a caramel wafer.
Armed with the solution the Architect shows on paper how to tie up all
of the values of place value 1 using a full adder.
This produces a value of place value 1 together with a carry-forward
of place value 2 thus allowing the Architect to use the same trick
again for the (now three) bits of place value 2.
They pause to reinforce how useful the carry-in is for
composition. They can then finish the problem showing the solution
given in figure \ref{fig:finDiag}.

With the finished solution the Architect pauses, asking the class if
they understand this solution.
Hopefully the class will air any questions they have, however, should
they fail to, the Technician can jump in to ask questions.
In the case where the Technician feels the Architect was unclear with
any part of their explanation they can offer an alternative
explanation, or asked questions which tease out more information.
When both of the lecturers feel comfortable that the class has a good
grasp on the problem, or at least that the students have no further questions,
and proposed solution the Architect asks the Technician to write the code
that represents this diagram.

\subsection{E for Encode}

\begin{figure}[h]
      \includegraphics[width=0.325\textwidth]{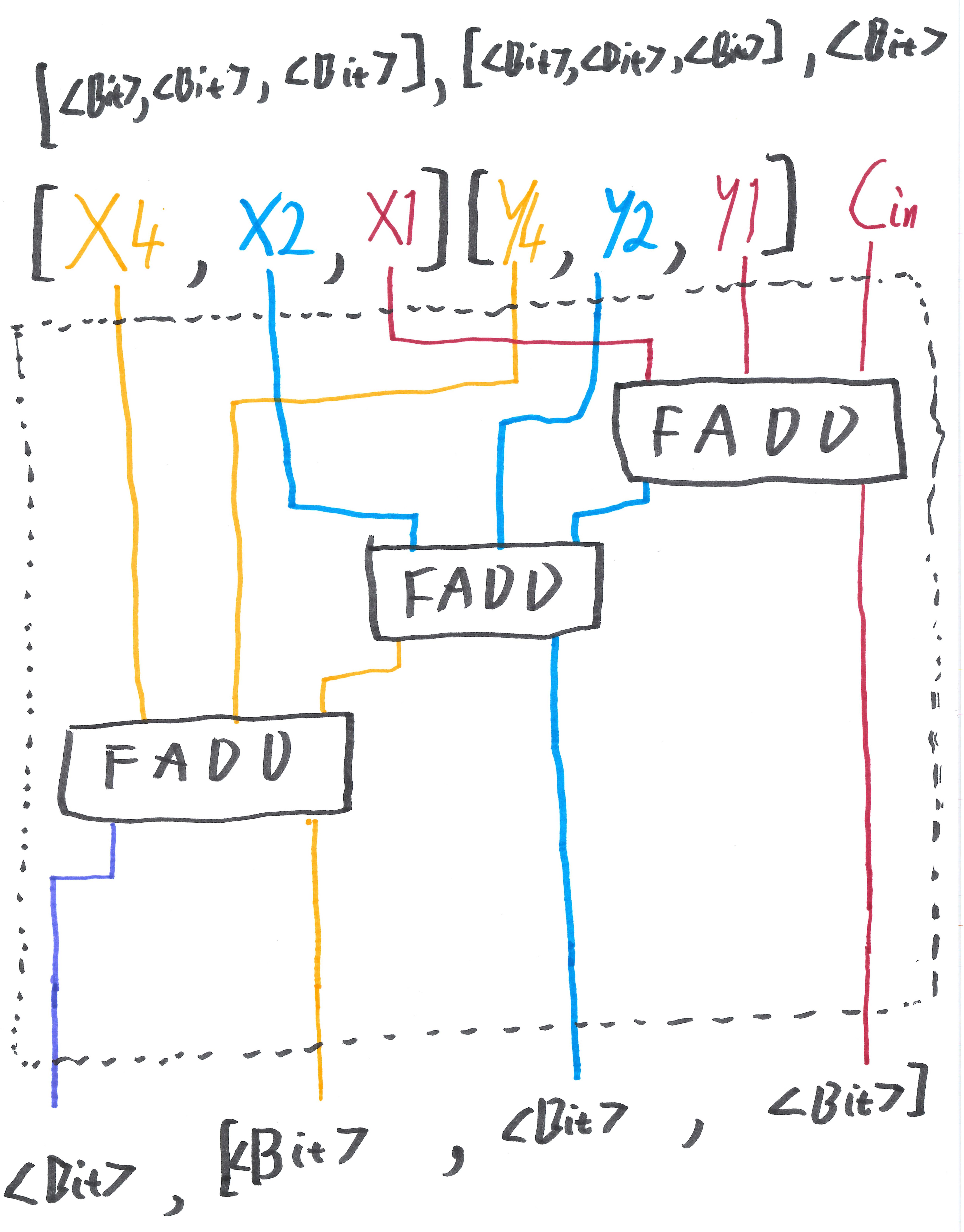}
      \caption{Input and Output Wires}
      \label{fig:IO}
\end{figure}

The Technician begins by explaining what we need to implement a Syrup
circuit.
First, a type signature specifying the circuit's interface.
The Technician takes a pen and draws a dotted box around the circuit,
being careful to make sure input and output wires overrun the box.
They then explain that this shows how the circuit interacts with the
outside world, or in other words what the circuit expects as inputs
and what it gives back as outputs.
The Technician then annotates each input and output with a type written
using Syrup syntax. In our running example, these annotations are respectively
cables of size 3
(\syrupCable{\syrupBit\syrupKeyword{,} \syrupBit\syrupKeyword{,} \syrupBit})
and single bits (\syrupBit) as seen in figure \ref{fig:IO}.

With the diagram updated, they move to the editor and start to define
the circuit.
They explain again that Syrup needs to know the inputs and outputs to
define a circuit, but happily they can simply read this from the
diagram!
Noting that repeatedly writing the type of cables of three bits is a
little bit tedious, they start by introducing a type
synonym \syrupType{Bit3} for it (line 1 below).
This has the added benefit of making the meaning of the circuit
clearer: an \syrupFunName{rca3} takes two 3 bit numbers and
a carry in and return a carry out and a 3 bit number (line 2).

\begin{syrup}
\syrupTypeDecl{Bit3}{\syrupCable{\syrupBit\syrupKeyword{,} \syrupBit\syrupKeyword{,} \syrupBit}}

\syrupFunDecl{rca3}
  {\syrupType{Bit3}\syrupKeyword{,} \syrupType{Bit3}\syrupKeyword{,} \syrupBit}
  {\syrupBit\syrupKeyword{,} \syrupType{Bit3}}

\end{syrup}

\begin{figure}[h]
      \includegraphics[width=0.325\textwidth]{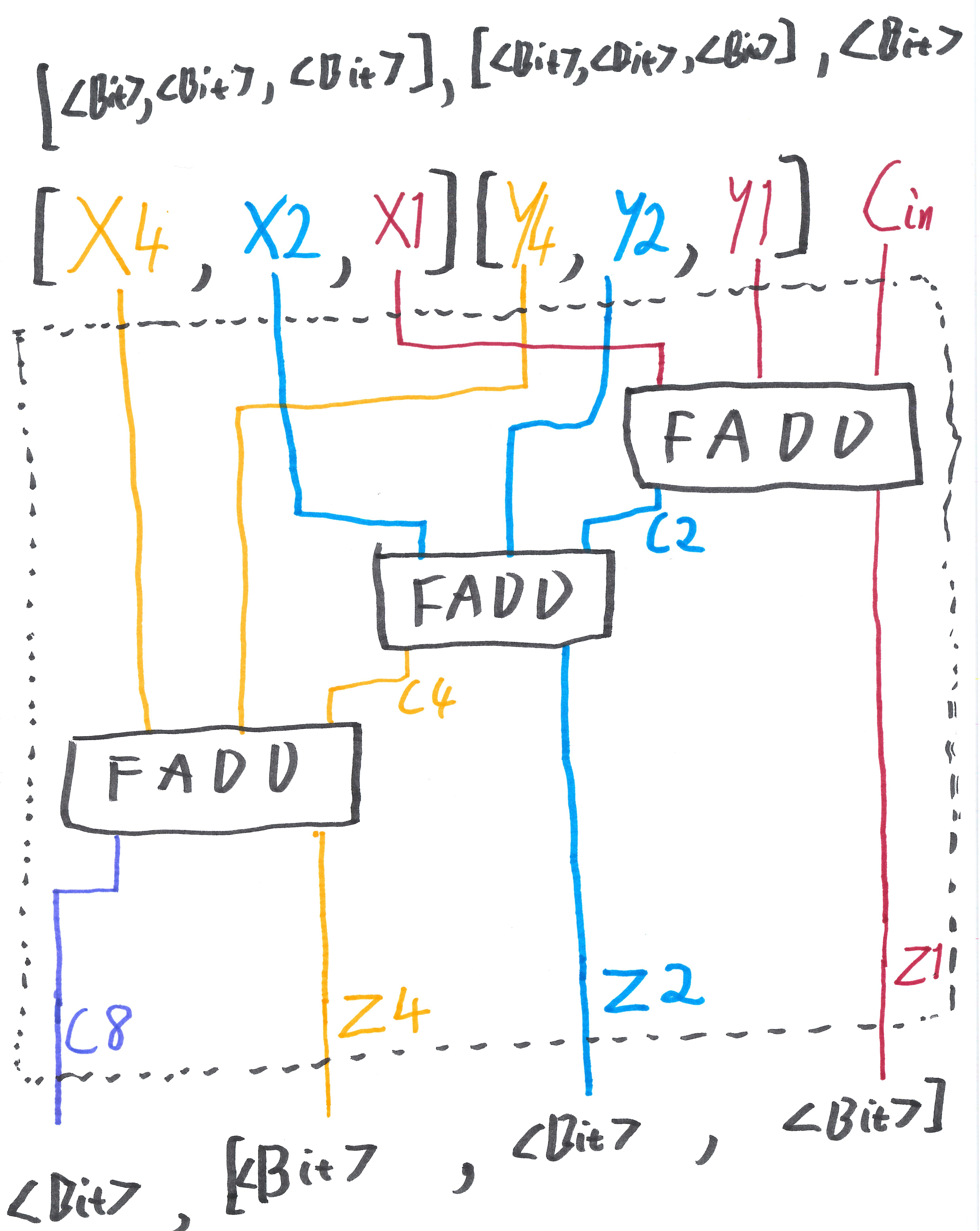}
      \caption{Intermediate Wire Naming and Output Values}
      \label{fig:inter}
\end{figure}

We now have the beginning of a circuit definition.
The Technician explains that we have told Syrup the type of our
circuit but not its implementation.
So the Technician returns to the diagram.
They explain to the class that we know how the circuit works at a high
level but it would be better to have small, easy to understand
sections that we can implement, rather than trying to write the whole
thing in one line.
Using the diagram they explain that it would be helpful to name all of
the outputs so that we can explain how to build them.
This helps, they explain, but we still have unnamed wires.
These wires correspond to intermediate results that are produced by
some gates and then fed into others.
They explain really these ought to have names so that we can tell
Syrup how to build them as well.
The Technician then labels the intermediate wires C2 and C4 to show
that they are carry-forwards into the 2s and 4s respectively, this
is shown in figure \ref{fig:inter}.

With the final labelled diagram in hand the Technician then begins
building the circuit.
They encode the outputs as a bit representing the a carry-forward to
the 8s, \syrupVarName{C8}, and the resulting three bit
sum packed as a cable of 3 bits
\syrupCable{\syrupVarName{Z4}\syrupKeyword{,}
\syrupVarName{Z2} \syrupKeyword{,} \syrupVarName{Z1}}.
They then show that we can explain how to compute these signals by
building small blocks at a time.
As with the draftsman's diagram they tie up the values of place value
1 using a full adder, labelling the output as \syrupVarName{Z1}, the
bit produced for output of place value 1 and the carry-forward into
place value 2 \syrupVarName{C2}.
In that process, they note that the name \syrupVarName{Cin} used in
the diagram breaks the mnemonic of labelling each name with its place
value and insist on using \syrupVarName{C1} instead.
The Architect sulks but agrees that the Technician is correct.
They carry on, using the labelled diagram to repeatedly ask ``How do I
define this'' and follow on by reading it from the diagram and
encoding it as Syrup code.
At each step they include a comment (\syrupComment{Using this syntax})
to document what the purpose of the intermediate computations is.

\begin{syrup}%



\syrupFunDefnWhereBlock{rca3}%
  {\syrupCable{\syrupVarName{X4}\syrupKeyword{,} \syrupVarName{X2}\syrupKeyword{,} \syrupVarName{X1}}\syrupKeyword{,} \syrupCable{\syrupVarName{Y4}\syrupKeyword{,} \syrupVarName{Y2}\syrupKeyword{,} \syrupVarName{Y1}}\syrupKeyword{,} C1}%
  {\syrupVarName{C8}\syrupKeyword{,} \syrupCable{\syrupVarName{Z4}\syrupKeyword{,} \syrupVarName{Z2}\syrupKeyword{,} \syrupVarName{Z1}}}
  {\syrupComment{Adding up the 1s} \\
   \syrupEquation{\syrupVarName{C2}\syrupKeyword{,} \syrupVarName{Z1}}{fadd}{\syrupVarName{X1}\syrupKeyword{,} \syrupVarName{Y1}\syrupKeyword{,} \syrupVarName{C1}} \\
   \syrupComment{Adding up the 2s} \\
   \syrupEquation{\syrupVarName{C4}\syrupKeyword{,} \syrupVarName{Z2}}{fadd}{\syrupVarName{X2}\syrupKeyword{,} \syrupVarName{Y2}\syrupKeyword{,} \syrupVarName{C2}} \\
   \syrupComment{Adding up the 4s} \\
   \syrupEquation{\syrupVarName{C8}\syrupKeyword{,} \syrupVarName{Z4}}{fadd}{\syrupVarName{X4}\syrupKeyword{,} \syrupVarName{Y4}\syrupKeyword{,} \syrupVarName{C4}} \\
  }

\end{syrup}

Internally, Syrup makes sure that the types are respected, that all
of the named wires are properly defined, that the definitions are not
circular in nature.
Once the code is accepted, we know we have a valid circuit.
But is it the one we wanted? The one we really wanted?

\subsection{D for Display}

\begin{figure}[h]
  \includegraphics[width=.275\textwidth]{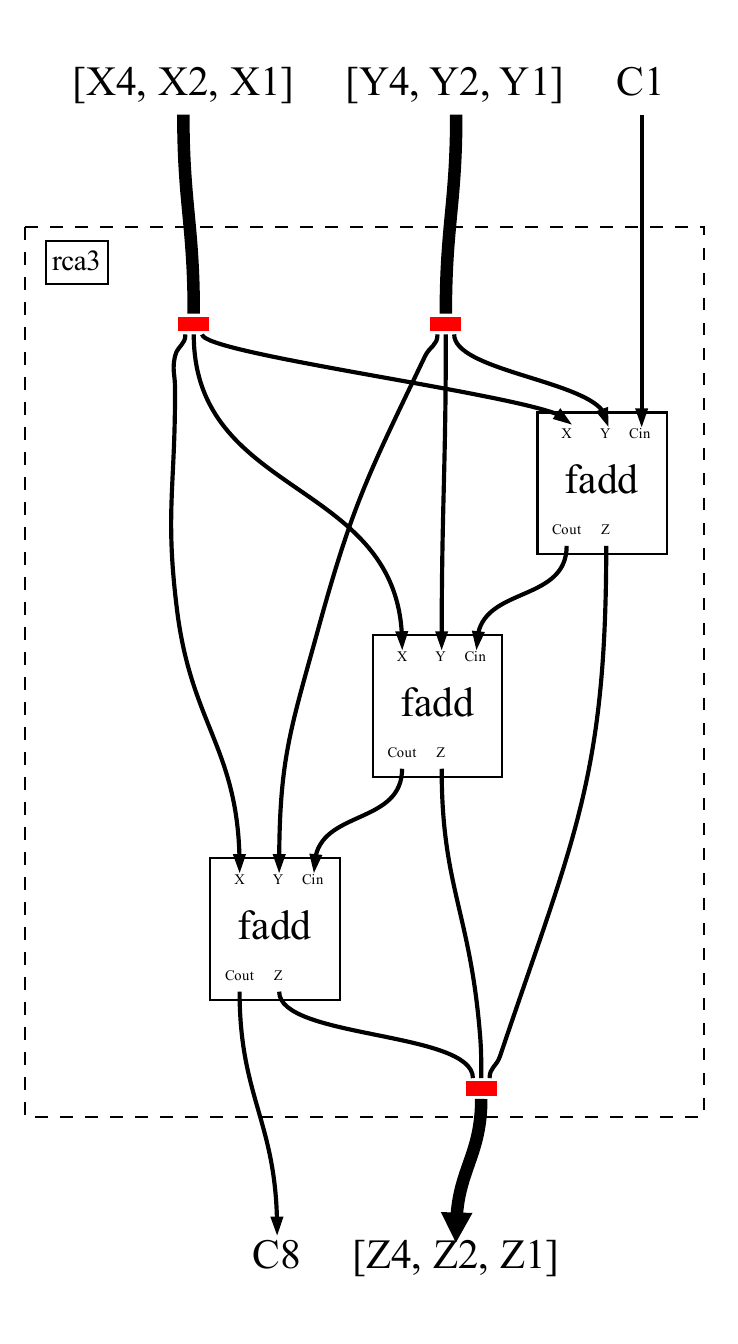}
  \caption{Displaying the \syrupFunName{rca3} circuit}
  \label{fig:syruprca}
\end{figure}

The Architect insists on using the display feature to obtain a circuit
diagram corresponding to the final piece of Syrup code.
The generated diagram, shown in figure \ref{fig:syruprca} can then be
compared to the original hand-drawn solution.
The lecturers then explain that both of the representations encode the
same logic and that if you have one you can generate the other.
This routine is performed regularly throughout the course and often
more than once during a lecture as solutions are refined or improved upon.

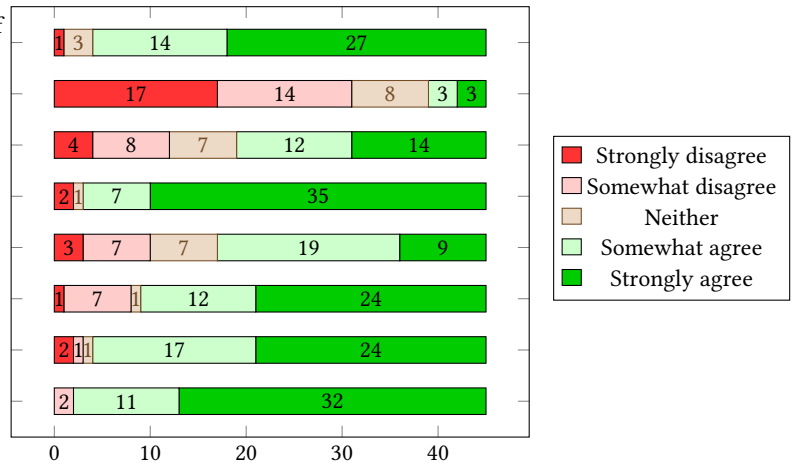
\begin{figure*}[h]

\pgfplotstableset{col sep=comma}

\begin{tikzpicture}
\begin{axis}[
xbar stacked,
stack negative=separate,
/pgf/number format/1000 sep=,
ytick distance=1,
ytick = {0,1,...,8},
yticklabels from table ={datasets/survey_results.csv}{Q},
y tick label style={align=left, text width=.415\textwidth},
nodes near coords,
legend style={
  at={(1.275,0.7)},
  anchor=north
  ,legend columns=1},
]
\addplot [fill=red!80!white] table [col sep=comma,x=Strongly disagree,y expr=\coordindex] {datasets/survey_results.csv};
\addplot [fill=red!20!white] table [col sep=comma,x=Somewhat disagree,y expr=\coordindex] {datasets/survey_results.csv};
\addplot table [col sep=comma,x=Neither agree nor disagree,y expr=\coordindex] {datasets/survey_results.csv};
\addplot [fill=green!20!white] table [col sep=comma,x=Somewhat agree,y expr=\coordindex] {datasets/survey_results.csv};
\addplot [fill=green!80!black] table [col sep=comma,x=Strongly agree,y expr=\coordindex] {datasets/survey_results.csv};
\legend{Strongly disagree,Somewhat disagree,Neither,Somewhat agree,Strongly agree}
\end{axis}
\end{tikzpicture}
\caption{Survey Results}
\label{table:Quant}
\end{figure*}

\section{Results and Analysis}

We aimed to understand whether or not students found the DED loop
helpful in their understanding.
In particular we wished to understand whether or not students prefered
this approach to code first.
To this end we designed a survey consistng of Likert scale questions.
The survey was distributed to first and second year students that had
attended semester 1 of \blinding{CS106} and therefore experienced the DED loop.
The start of the survey contained a description of the DED loop and
its deployment.
The survey was distributed via email and students were given time
during class to complete the survey.
Participation was entirely voluntary and ethical consent was granted by
\blinding{the University of Strathclyde Department of Computer and
Information Sciences Ethics board}.

After dissemination 45 complete responses were obtained, the responses to the Likert scale questions are displayed in figure \ref{table:Quant}.
Responses were highly positive with 43 of the respondents agreeing that the approach was enging, 41 respondents agreeing that the DED loop made it easier to follow lectures and 36 agreeing that the DED loop made it easier to understand Syrup.
Less students, though still a majority of 28 responses, agreed to some extent that they used the DED loop when writing their own code.
A large majority of students disagreed with the statement ``I would have prefered if only Syrup was used''. Students were also very positive about the Co-Lecturing approach with 42 of the 45 responses agreeing that two lecturers presenting differing points of view was helpful.
41 of the students agreed that the use of colours was helpful.

Students were overwhelmingly positive about the DED approach to circuit specification.
In particular students were heavily in favour of this deployment of Co-Lecturing with 42 of the students agreeing that mutiple perspectives were helpful.
This directly answers RQ3, students did find co-lecturing helpful.
A majority of 31 students disagreed that they would have prefered if only Syrup was used, 6 students agreed that they would have prefered a Syrup only approach.
These results address RQ2 students prefer the DED loop to a code-only approach.
Taken together our results answer RQ1, students did find the DED loop helpful.


While our preliminary results were very positive, we do acknowledge some limitations to our work.
Our questions focussed on whether or not students found the approach helpful. As this approach was integral to our delivery of the course it would have been very difficult to tease out whether or not student understanding had been improved outside of students perceived understanding. Ideally a controlled experiment would have been used to determine the efficacy of the DED loop, as such our results are preliminary.

Ideally our survey would have had more responses. Given the number of responses it is likely that selection bias has played a role in the positive nature of our responses and that a larger survey may have shown a larger variety of opinions.

\section{Conclusion and Future Work}

We present evidence that students are highly positive about the DED loop approach to circuit design education. Our results show that the DED loop has merit and see areas for future work.

It would be interesting to understand whether or not students are more capable at generating the circuit diagram from a Syrup specification or vice versa. The DED loops focuses on solving via a diagram and then deriving the code. It would be interesting to understand if this forms a bidirectional understanding or simply a one way derivation technique.

Students were also very positive about the co-lecturing aspect of the DED loop. It would be interesting to see whether this approach functions as well with only one lecturer. It may be that students find it less engaging when the two lecturers cannot discuss ideas but similarly helpful in understanding.

Our work introduces Syrup, a pedagogy focused hardware description language.
We also introduce the DED loop, a lecturing dynamic which helps students to bridge the gap from intuitive circuit diagrams to hardware specification code.
Our approach is simple, easy to reproduce and, as our data shows, popular with students.
We encourage the reader to try this approach themselves, it's DED easy.

\clearpage
\onecolumn \begin{multicols}{2}

\bibliographystyle{ACM-Reference-Format}
\bibliography{DED}

@inproceedings{Lambert2025,
author = {Lambert, Alasdair and Gale, Stuart and Schoen, Ezra},
title = {Double Acts for Demystifying Subroutine Calling Conventions},
year = {2025},
isbn = {9798400715693},
publisher = {Association for Computing Machinery},
address = {New York, NY, USA},
url = {https://doi.org/10.1145/3724389.3731255},
doi = {10.1145/3724389.3731255},
booktitle = {Proceedings of the 30th ACM Conference on Innovation and Technology in Computer Science Education V. 2},
pages = {733–734},
numpages = {2},
location = {Nijmegen, Netherlands},
series = {ITiCSE 2025}
}

@inproceedings{Fagan2025,
author = {Fagan, Andrew and Lambert, Alasdair and Goodfellow, Martin},
title = {Learning Programming Languages by Pantomime},
year = {2025},
isbn = {9798400711725},
publisher = {Association for Computing Machinery},
address = {New York, NY, USA},
url = {https://doi.org/10.1145/3702212.3702213},
doi = {10.1145/3702212.3702213},
booktitle = {Proceedings of the 9th Conference on Computing Education Practice},
pages = {1–4},
numpages = {4},
location = {
},
series = {CEP '25}
}

@article{Baeten2014,
   author = {Marlies Baeten and Mathea Simons},
   doi = {10.1016/J.TATE.2014.03.010},
   issn = {0742-051X},
   journal = {Teaching and Teacher Education},
   month = {7},
   pages = {92-110},
   publisher = {Pergamon},
   title = {Student teachers' team teaching: Models, effects, and conditions for implementation},
   volume = {41},
   year = {2014},
}

@article{Zehetmeier2018,
   author = {Daniela Zehetmeier and Axel Böttcher and Anne Brüggemann-Klein},
   doi = {10.4995/HEAD18.2018.8103},
   isbn = {9788490486900},
   issn = {2603-5871},
   journal = {4th International Conference on Higher Education Advances (HEAD'18)},
   month = {7},
   pages = {873-880},
   publisher = {Editorial Universitat Politècnica de València},
   title = {Designing Lectures as a Team and Teaching in Pairs},
   url = {https://riunet.upv.es/handle/10251/109639},
   year = {2018},
}

@article{Liebel2017,
   author = {Grischa Liebel and Håkan Burden and Rogardt Heldal},
   doi = {10.1080/13562517.2016.1221811},
   issn = {14701294},
   issue = {1},
   journal = {Teaching in Higher Education},
   month = {1},
   pages = {62-77},
   publisher = {Routledge},
   title = {For free: continuity and change by team teaching},
   volume = {22},
   url = {https://www.tandfonline.com/doi/abs/10.1080/13562517.2016.1221811},
   year = {2017},
}

@inproceedings{coTA,
author = {Patitsas, Elizabeth},
title = {A case study of the development of CS teaching assistants and their experiences with team teaching},
year = {2013},
isbn = {9781450324823},
publisher = {Association for Computing Machinery},
address = {New York, NY, USA},
url = {https://doi.org/10.1145/2526968.2526981},
doi = {10.1145/2526968.2526981},
booktitle = {Proceedings of the 13th Koli Calling International Conference on Computing Education Research},
pages = {115–124},
numpages = {10},
location = {Koli, Finland},
series = {Koli Calling '13}
}

@article{costa1993through,
  title={Through the lens of a critical friend},
  author={Costa, Arthur L and Kallick, Bena and others},
  journal={Educational leadership},
  volume={51},
  pages={49--49},
  year={1993},
  publisher={ASCD ASSOCIATION FOR SUPERVISION AND}
}

@inproceedings{caReview,
author = {Shah, Anshul and Soosai Raj, Adalbert Gerald},
title = {A Review of Cognitive Apprenticeship Methods in Computing Education Research},
year = {2024},
isbn = {9798400704239},
publisher = {Association for Computing Machinery},
address = {New York, NY, USA},
url = {https://doi.org/10.1145/3626252.3630769},
doi = {10.1145/3626252.3630769},
booktitle = {Proceedings of the 55th ACM Technical Symposium on Computer Science Education V. 1},
pages = {1202–1208},
numpages = {7},
location = {Portland, OR, USA},
series = {SIGCSE 2024}
}

@inproceedings{liveEval,
author = {Shah, Anshul and Hogan, Emma and Agarwal, Vardhan and Driscoll, John and Porter, Leo and Griswold, William G. and Soosai Raj, Adalbert Gerald},
title = {An Empirical Evaluation of Live Coding in CS1},
year = {2023},
isbn = {9781450399760},
publisher = {Association for Computing Machinery},
address = {New York, NY, USA},
url = {https://doi.org/10.1145/3568813.3600122},
doi = {10.1145/3568813.3600122},
booktitle = {Proceedings of the 2023 ACM Conference on International Computing Education Research - Volume 1},
pages = {476–494},
numpages = {19},
location = {Chicago, IL, USA},
series = {ICER '23}
}

@article{Knobelsdorf2014,
   author = {Maria Knobelsdorf and Christoph Kreitz and Sebastian Böhne},
   doi = {10.1145/2538862.2538944},
   isbn = {9781450326056},
   journal = {SIGCSE 2014 - Proceedings of the 45th ACM Technical Symposium on Computer Science Education},
   pages = {67-72},
   publisher = {Association for Computing Machinery},
   title = {Teaching theoretical computer science using a cognitive apprenticeship approach},
   url = {https://dl.acm.org/doi/10.1145/2538862.2538944},
   year = {2014},
}

@article{Yu2024,
   author = {Chih-Chang Yu and Leon Yufeng Wu},
   issn = {11763647, 14364522},
   issue = {2},
   journal = {Educational Technology \& Society},
   pages = {pp. 183–196},
   publisher = {International Forum of Educational Technology \& Society, National Taiwan Normal University, Taiwan},
   title = {Instructing with Cognitive Apprenticeship Programming Learning System (CAPLS) for novice computer science college freshmen: An exploration study},
   volume = {27},
   url = {https://www.jstor.org/stable/48766170},
   year = {2024},
}

@article{paxton2002,
author = {Paxton, John},
title = {Live programming as a lecture technique},
year = {2002},
issue_date = {December 2002},
publisher = {Consortium for Computing Sciences in Colleges},
address = {Evansville, IN, USA},
volume = {18},
number = {2},
issn = {1937-4771},
journal = {J. Comput. Sci. Coll.},
month = {dec},
pages = {51–56},
numpages = {6}
}

@article{collins1991cognitive,
  title={Cognitive apprenticeship: Making thinking visible},
  author={Collins, Allan and Brown, John Seely and Holum, Ann and others},
  journal={American educator},
  volume={15},
  number={3},
  pages={6--11},
  year={1991}
}

@inproceedings{Selvaraj2021,
author = {Selvaraj, Ana and Zhang, Eda and Porter, Leo and Soosai Raj, Adalbert Gerald},
title = {Live Coding: A Review of the Literature},
year = {2021},
isbn = {9781450382144},
publisher = {Association for Computing Machinery},
address = {New York, NY, USA},
url = {https://doi.org/10.1145/3430665.3456382},
doi = {10.1145/3430665.3456382},
booktitle = {Proceedings of the 26th ACM Conference on Innovation and Technology in Computer Science Education V. 1},
pages = {164–170},
numpages = {7},
location = {Virtual Event, Germany},
series = {ITiCSE '21}
}

@inproceedings{raj2018role,
  title={Role of live-coding in learning introductory programming},
  author={Raj, Adalbert Gerald Soosai and Patel, Jignesh M and Halverson, Richard and Halverson, Erica Rosenfeld},
  booktitle={Proceedings of the 18th koli calling international conference on computing education research},
  pages={1--8},
  year={2018}
}

@inproceedings{schoolBlock,
author = {Weintrop, David and Wilensky, Uri},
title = {To block or not to block, that is the question: students' perceptions of blocks-based programming},
year = {2015},
isbn = {9781450335904},
publisher = {Association for Computing Machinery},
address = {New York, NY, USA},
url = {https://doi.org/10.1145/2771839.2771860},
doi = {10.1145/2771839.2771860},
booktitle = {Proceedings of the 14th International Conference on Interaction Design and Children},
pages = {199–208},
numpages = {10},
location = {Boston, Massachusetts},
series = {IDC '15}
}

@article{blockCS,
author = {Weintrop, David},
title = {Block-based programming in computer science education},
year = {2019},
issue_date = {August 2019},
publisher = {Association for Computing Machinery},
address = {New York, NY, USA},
volume = {62},
number = {8},
issn = {0001-0782},
url = {https://doi.org/10.1145/3341221},
doi = {10.1145/3341221},
journal = {Commun. ACM},
month = jul,
pages = {22–25},
numpages = {4}
}

@article{blockvstext,
author = {Weintrop, David and Wilensky, Uri},
title = {Comparing Block-Based and Text-Based Programming in High School Computer Science Classrooms},
year = {2017},
issue_date = {March 2018},
publisher = {Association for Computing Machinery},
address = {New York, NY, USA},
volume = {18},
number = {1},
url = {https://doi.org/10.1145/3089799},
doi = {10.1145/3089799},
journal = {ACM Trans. Comput. Educ.},
month = oct,
articleno = {3},
numpages = {25}
}

@INPROCEEDINGS{transition2text,
  author={Moors, Luke and Luxton-Reilly, Andrew and Denny, Paul},
  booktitle={2018 International Conference on Learning and Teaching in Computing and Engineering (LaTICE)},
  title={Transitioning from Block-Based to Text-Based Programming Languages},
  year={2018},
  volume={},
  number={},
  pages={57-64},
  doi={10.1109/LaTICE.2018.000-5}
  }

@article{block2textMeta,
title = {Supporting learners in the transition from block-based to text-based programming, a systematic review},
journal = {Journal of Computer Languages},
volume = {84},
pages = {101342},
year = {2025},
issn = {2590-1184},
doi = {https://doi.org/10.1016/j.cola.2025.101342},
url = {https://www.sciencedirect.com/science/article/pii/S2590118425000280},
author = {Glenn Strong and Nina Bresnihan and Brendan Tangney}
}

@article{blockornot,
author = {Tarık Talan and Yunus Doğan and Yusuf Kalinkara and Veli Batdi},
title = {To block or not to block? : a meta-analysis of effectiveness of block-based programming},
journal = {Research in Science and Technological Education},
volume = {0},
number = {0},
pages = {1--20},
year = {2025},
publisher = {Routledge},
doi = {10.1080/02635143.2025.2593936},


URL = {

        https://doi.org/10.1080/02635143.2025.2593936



},
eprint = {

        https://doi.org/10.1080/02635143.2025.2593936



}

}

@inproceedings{notional,
author = {Fincher, Sally and Jeuring, Johan and Miller, Craig S. and Donaldson, Peter and du Boulay, Benedict and Hauswirth, Matthias and Hellas, Arto and Hermans, Felienne and Lewis, Colleen and M\"{u}hling, Andreas and Pearce, Janice L. and Petersen, Andrew},
title = {Notional Machines in Computing Education: The Education of Attention},
year = {2020},
isbn = {9781450382939},
publisher = {Association for Computing Machinery},
address = {New York, NY, USA},
url = {https://doi.org/10.1145/3437800.3439202},
doi = {10.1145/3437800.3439202},
booktitle = {Proceedings of the Working Group Reports on Innovation and Technology in Computer Science Education},
pages = {21–50},
numpages = {30},
location = {Trondheim, Norway},
series = {ITiCSE-WGR '20}
}

\end{multicols}

\end{document}